\documentclass[11pt,twoside]{article}

\usepackage{asp2004}
\usepackage{graphicx}
\usepackage{natbib}

\begin{document}
\title{Understanding white dwarf binary evolution with white dwarf/main 
sequence binaries: first results from SEGUE}   
\author{M.R. Schreiber$^{1,2}$, A. Nebot Gomez-Moran$^2$, A.D. Schwope$^2$}   
\affil{$^1$ Universidad de Valparaiso, Facultad de Ciencias, Departamento de Fisica
y Meteorologia, Av. Gran Breta\~{n}a 1111, Valparaiso, Chile\\
$^2$ Astrophysikalisches Inst. Potsdam, An der Sternwarte 16, 14482 Potsdam,
Germany}

\begin{abstract} 
Close white dwarf binaries make up a wide variety of
objects such as double white dwarf binaries, which are possible 
SN Ia progenitors, cataclysmic variables, super soft sources, or AM CVn stars.
The evolution and formation of close white dwarf 
binaries crucially depends on the rate at which angular momentum is extracted
from the binary orbit. The two most important sources of angular
momentum loss are the common envelope phase and magnetic
braking. Both processes are so far poorly understood. Observational
population studies of white dwarf/main sequence binaries provide
the potential to significantly progress with this situation and to
clearly constrain magnetic braking and the CE-phase. However, the
current population of white dwarf/main sequence binaries is highly 
incomplete and heavily
biased towards young systems containing hot white dwarfs. 
The SDSSII/SEGUE collaboration awarded
us with 5 fibers per plate pair in order to fill this gap and to identify the
required unbiased sample of old white dwarf/main sequence binaries.
The success rate of our selection criteria exceeds $65\%$ and during the first 
10 months we have identified 41 new systems, most of them belonging
to the missed old population. 
\end{abstract}

\section{Introduction}

Close binaries containing at least one white dwarf span a wide range of
interesting and exotic stars, such as detached white dwarf binaries,
cataclysmic variables (CVs), or AM CVn binaries. 
Besides offering the opportunity to study physics under
extreme conditions, these objects are extremely important in the
general astrophysical context: Supernova\,Ia arise either from merging
binary white dwarfs or from interacting white dwarf/main sequence
binaries and AM CVn stars are expected to significantly contribute to the
gravitational wave background which will be measured by LISA. 
All the different types of close white dwarf binaries have two points 
in common: (1) they evolved through at
least one common envelope (CE) phase and (2) they undergo subsequent
orbital angular momentum loss (AML).  Sad but true, the physics of
both the CE and AML are very poorly understood. 

In current theories the CE phase is simply approximated
by a parameterized energy 
\citep{paczynski76-1,webbink84-1,willems+kolb04-1}
or angular momentum equation \citep{nelemans+tout05-1}.
Both descriptions differ significantly in the predicted outcome of the
CE phase and in both prescriptions the efficiency to
``use'' the orbital energy (angular momentum) to expel the envelope is 
very uncertain. Hence, the CE phase is probably the least understood period 
of close binary evolution.   
Once the envelope is expelled, the
evolution of the post common envelope binary (PCEB) is mainly driven
by AML due to magnetic braking.  Unfortunately, the two currently 
favoured prescriptions for magnetic braking 
\citep{verbunt+zwaan81-1,andronovetal03-1}, 
differ by up to two orders of magnitude. 
Even worse, it is not clear whether magnetic
braking is continuously present or if it gets disrupted when the
secondary star is fully convective. In order to explain the orbital period gap
observed in the period distribution of CVs,
one needs to assume the latter \citep[e.g.][]{king88-1,howelletal01-1} 
while observations of single low mass 
stars do not show any evidence for such a discontinuity
\citep[e.g.][]{pinsonneaultetal02-1}.  

Significant progress in the theoretical modelling of the CE phase 
and AML due to magnetic braking will clearly need observational input.  
A quantitative test of the current theories
requires the knowledge of a large and unbiased population
of close binaries that underwent a CE and subsequent orbital AML. 
The ideal class of stars to provide such observational
constraints on the CE and magnetic braking models are detached PCEBs 
consisting of a white dwarf and a main sequence
star, as (1) white dwarf binaries are intrinsically numerous, (2) the 
properties of both stellar components are well-understood, and (3) they have 
rather short orbital periods ($\sim\,2$h$-50$\,d).

\section{PCEBs in the pre-SDSS era}

\citet{schreiber+gaensicke03-1} analysed the population of PCEBs with 
determined orbital period and white dwarf temperature. Their sample consisted
of only 30 systems -- a surprisingly small number when compared with the more
than 1000 CVs listed in \citet{downesetal06-1}. Even worse, the detailed 
analysis of \citet{schreiber+gaensicke03-1} showed that the small sample
of 30 PCEBs is also heavily biased towards hot white dwarfs and late type
secondary star spectral types. This bias is a natural consequence of the way
PCEBs have been discovered in the past: as white dwarfs in the first place,
with some evidence for a faint red companion found later. Finally,
\citet{schreiber+gaensicke03-1} calculated the evolutionary time scale of the
30 young (containing hot white dwarfs) PCEBs and find that most of them have 
passed only a very small fraction of their PCEB lifetime. This immediately
leads to the prediction of a large population of old PCEBs containing cold 
white dwarfs which has not yet been identified. 

\section{The biases of the SDSS DR4 sample}

Since the first data release of the Sloan Digital Sky Survey (SDSS), 
the situation changed drastically. Based on SDSS imaging 
and some DR\,1 spectra \citet{smolcicetal04-1} identified 
a new stellar locus, i.e. the white dwarf/main sequence (WD/MS) 
binary bridge. 
The population of these WD/MS binaries consists 
of wide binaries that will never interact and whose components evolve 
like single stars and close binaries that went through a common envelope 
phase (PCEBs). 

The SDSS turned out to be also very efficient 
in spectroscopically identifying new unresolved WD/MS binaries.
Recently \citet{silvestrietal06-1} published a
list of $\sim\,747$ new WD/MS binary systems found in SDSS/DR4. 
However, as stated by \citet{silvestrietal06-1} themselves, the SDSS DR4 
sample is again subject to strong observational biases. 
The WD/MS systems identified in SDSS/DR4 originate primarily from two 
different channels: the colour selection described in 
\citet{silvestrietal06-1} and serendipitous objects from QSO fibres. 
The color selection used by \citet{silvestrietal06-1} selects
hot systems mainly because of the cut used in $u-g$ versus $g-r$  
and the SDSS QSO selection
algorithm \citep[see][Fig.\,7]{richardsetal02-1}
explicitly excludes the color-color space of cold white dwarf/main sequence
binaries. Hence both channels produce 
predominantly WD/MS binaries with hot white dwarfs, i.e. young objects, 
which -- according to \citet{schreiber+gaensicke03-1} --  represent only the 
minority of all WD/MS binaries. 

\section{Identifying old PCEBs with SEGUE}

A true constraint on AML mechanisms in close
binaries will only be possible once a {\em{representative}} sample of PCEBs 
has been identified. 
As partners of SDSS\,II we are 
running a successful program (PI: M. Schreiber) 
identifying the missing 
cold WD/MS binary population. The
SEGUE-collaboration awarded us with 5 fibers per SEGUE plate pair
($\sim\,7$deg$^2$) and we developed special color-cuts to 
select WD/MS systems containing cold white dwarfs, i.e. 
\begin{eqnarray} 
u-g & < & 2.25	\hspace{1cm}g-r  >  -19.78*(r-i) + 11.13	\nonumber\\
g-r & > & -0.2 \hspace{1cm}g-r  <  0.95*(r-i) + 0.5\nonumber\\
g-r & < & 1.2\hspace{1cm}i-z  >  0.5 \hspace{0.2cm}\mathrm{for}\hspace{0.2cm}r-i > 1.0\nonumber\\
r-i & > & 0.5\hspace{1cm}i-z  >  0.68*(r-i)-0.18 \hspace{0.2cm}\mathrm{for} \hspace{0.2cm}r-i <= 1.0\nonumber\\
r-i & < & 2.0\hspace{1cm}15 <  g  < 20.\nonumber
\end{eqnarray}
The main selection criteria are shown in Fig.\,1 as black lines. 
In the first 10 months 41 SEGUE-plates with WD/MS target selection have been
observed. During the first drilling run in Oct.\,2005, the above criteria 
have been applied to reddening corrected magnitudes. This led to the
identification of several nearby single M-dwarfs whose reddening corrected
colors resemble WD/MS binaries containing cold white dwarfs. 
The success rate for the Oct.\,2005 plates therefore is only $14/35=40\%$
on 22 plates.  
Since 2006 we use non-corrected $ugriz$ magnitudes and our success rate
increased to $27/40=67.5\%$ on 19 plates. 
Fig.\,1 shows the positions of the 75 SEGUE-WD/MS candidates including the 41 
WD/MS systems (black open squares). Also shown are the 
\citet{silvestrietal06-1} sample (black points) and the QSO and single star 
population (gray). As an example for the 41 identified systems, Fig.\,2 shows
the SEGUE spectrum of one cool WD/MS binary.
\begin{figure}[!th]
\begin{center}
\includegraphics[angle=270,,clip=,width=12cm]{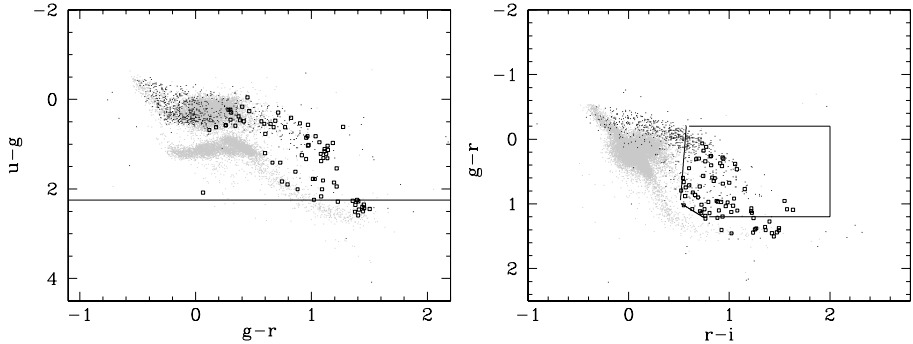}
\end{center}
\caption[]{The SEGUE-WD/MS color cuts (black lines) in two color-color
diagrams. Quasars and single stars are shown in grey. 
The (not reddening corrected) positions of the 75 WD/MS candidates 
selected during the first 10 months are marked as open squares. In the first 
drilling run we used reddening corrected
magnitudes and some nearby M-dwarfs (those below the lower vertical lines) 
appeared as WD/MS candidates. 
Since 2006 we select our candidates without reddening correction and the 
success rate increased to $67.5\%$. 
The SDSS/DR4 WD/MS population
\citep{silvestrietal06-1} is shown as black points. Apparently, the overlap of
the with the SEGUE selection is rather small as the latter is especially
designed to identify the missing old population.}
\end{figure}
\begin{figure}[!th]
\begin{center}
\includegraphics[angle=270,,clip=,width=11.5cm]{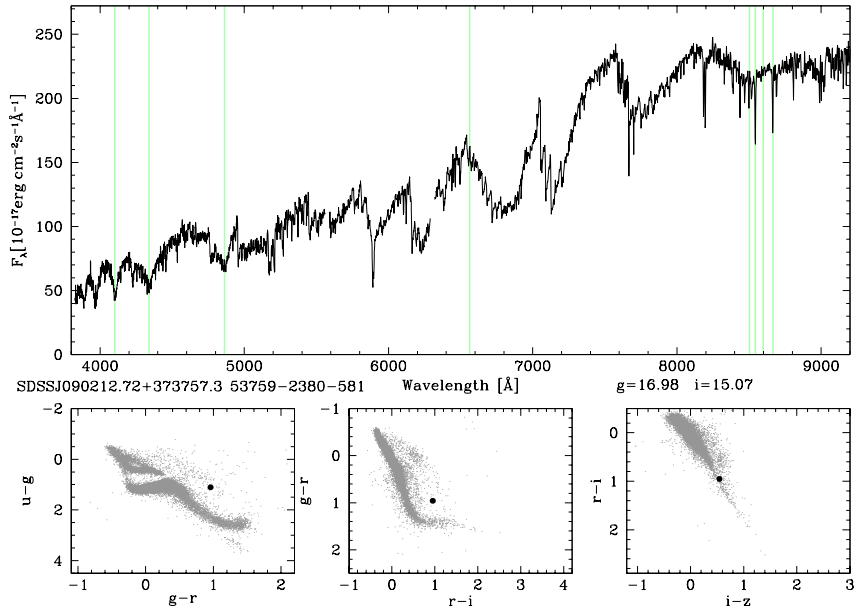}
\end{center}
\caption[]{The spectrum of a SEGUE WD/MS binary and the position in
color-color space (using fiber-magnitudes).   
}
\end{figure}

We determined the white dwarf temperature and the spectral type of the
secondary of the 41 WD/MS binaries by fitting simultaneously the 
composite binary spectrum. The resulting distributions are shown in Fig.\,3. 
Compared to the SDSS DR4 sample published by \citet{silvestrietal06-1} our
sample contains significantly more WD/MS systems with cold white dwarfs
and/or early type secondary stars thereby overcoming previous biases in the
sample of known WD/MS binaries.

\begin{figure}[!th]
\begin{center}
\includegraphics[angle=270,,clip=,width=9cm]{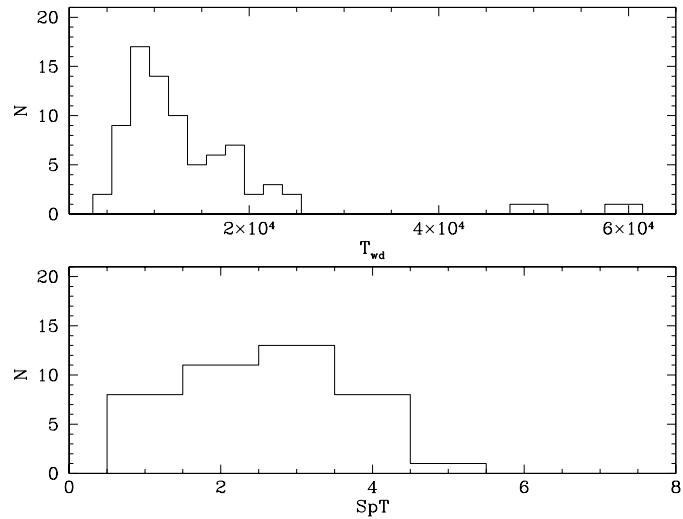}
\end{center}
\caption[]{The distributions of the effective temperature of the white dwarf
(top) and spectral type of the main sequence secondary star
(bottom). As predicted, compared to \citet[][their Fig.\,5 and
10]{silvestrietal06-1} the SEGUE-WD/MS sample contains significantly more 
systems with cold white dwarfs and early type secondaries. }
\end{figure}

According to the SEGUE baseline $\sim200$ plate pairs will be observed until
mid 2008 and we therefore expect to identify $\sim\,400-500$ new WD/MS 
binaries by then.
Together with the systems identified by 
\citet{silvestrietal06-1} the resulting more than $\sim1200$ WD/MS systems 
will form {\em{the}} data base to constrain theories of close binary 
evolution.

\section{Constraining close binary evolution with PCEBs}

In principal the three big questions of close binary evolution 
can be answered using a large sample of PCEBs with known orbital period,
secondary spectral type, and white dwarf temperature:
{\bf{(1)}} The disrupted magnetic braking scenario predicts an increase of
the relative number of PCEBs by a factor $\sim\,1.7$ in the range of
secondary spectral types M3-M5 \citep[see][]{politano+weiler06-1}.  
To confirm or disprove the predicted increase one needs to identify 
PCEBs with M3-M5 secondaries among the WD/MS population.  
As the mean PCEB lifetime can be rather large, the expected increase of the
relative number of PCEBs will be more pronounced in the old SEGUE population. 
{\bf{(2)}} The strength of AML can be estimated by comparing the orbital
period distributions of PCEBs at different times of the PCEB
evolution. A representative sample of PCEBs for
secondary spectral types M0-M8 and effective temperatures of the white
dwarf of $T_{\mathrm{wd}}\sim\,10000-40000\,$K is required. 
{\bf{(3)}} The predictions
of the two currently favoured prescriptions of the CE phase differ in
particular in the predicted orbital period distribution of long orbital 
period PCEBs \citep{nelemans+tout05-1}. 
Consequently, identifying the 
long orbital period end of the PCEB population will clearly constrain current
theories of the CE phase. 

To sum up, characterizing a large sample of PCEBs provides the potential 
to solve the three most important problems in close
binary evolution. To that end we have initiated a large-scale follow-up
programme to identify and characterize the PCEBs among the WD/MS sample
involving telescopes at both hemispheres and
utilizing multi-epoch spectroscopy, time-resolved photometry, and astrometry 
with promising first results.


\acknowledgements 
We acknowledge support by the Deutsches Zentrum f\"ur Luft- und Raumfahrt
(DLR) under contract FKZ\,50\,OR\,0404 (MRS, ANGM) and FONDECYT, grant 1061199 
(MRS). 
Funding for the SDSS and SDSS-II has been provided by the A. P. Sloan
Foundation, the Participating Institutions, the NSF,
the U.S. Dept. of Energy, the Nat. Aeronautics and Space
Admin., the Japanese Monbukagakusho, the Max Planck Society, and the
Higher Education Funding Council for England. The SDSS Web Site is
http://www.sdss.org/. 

The SDSS is managed by the Astrophysical Research Consortium for the
Participating Institutions. The Participating Institutions are the American
Museum of Nat. History, Astrophys. Inst. Potsdam, University of
Basel, University of Cambridge, Case Western Reserve University, University of
Chicago, Drexel University, Fermilab, the Institute for Advanced Study, the
Japan Participation Group, Johns Hopkins University, the Joint Institute for
Nuclear Astrophysics, the Kavli Institute for Particle Astrophysics and
Cosmology, the Korean Scientist Group, the Chinese Academy of Sciences
(LAMOST), Los Alamos National Laboratory, the Max-Planck-Institute for
Astronomy (MPIA), the Max-Planck-Institute for Astrophysics (MPA), New Mexico
State University, Ohio State University, University of Pittsburgh, University
of Portsmouth, Princeton University, the United States Naval Observatory, and
the University of Washington. 




\end{document}